\documentclass[12pt]{iopart}

\usepackage{amssymb}
\usepackage{color}
\usepackage{graphicx}

\usepackage{cite}

\newcommand{\eqref}[1]{(\ref{#1})}

\newcommand{\JU}{J/U}
\newcommand{\Jcrit}{\big(\JU\big)_{\rm c}}

\newcommand{\muU}{\mu/U}

\newcommand{\psimin}{\psi_{\min}}

\newcommand{\betacrit}{\beta_{\rm crit}}
\begin{document}

\title[Critical exponents of the Bose-Hubbard model]
      {Hypergeometric continuation of divergent perturbation series. \\
      I. Critical exponents of the Bose-Hubbard model}

\author{S\"oren Sanders$^1$, Martin Holthaus$^1$}

\address{$^1$ Institut f\"ur Physik, Carl von Ossietzky Universit\"at, D-26111 Oldenburg, Germany}

\ead{soeren.sanders@uni-oldenburg.de}

\date{\today}

\begin{abstract}
We study the connection between the exponent of the order parameter of the Mott insulator-to-superfluid transition occurring in the two-dimensional Bose-Hubbard model, and the divergence exponents of its one- and two-particle correlation functions. We find that at the multicritical points all divergence exponents are related to each other, allowing us to express the critical exponent in terms of one single divergence exponent. This approach correctly reproduces the critical exponent of the three-dimensional $XY$ universality class. Because divergence exponents can be computed in an efficient manner by hypergeometric analytic continuation, our strategy is applicable to a wide class of systems.
\end{abstract} 

% \keywords{quantum phase transition, critical exponents, Bose-Hubbard model, analytic continuation}


% 05.30.Rt	Quantum phase transitions
% 02.30.Mv	Approximations and expansions
% 11.15.Bt	General properties of perturbation theory
% 05.30.Jp	Boson systems

\maketitle 

%%%%%%%%%%%%%%%%%%%%%%%%%%%%%%%%%%%%%%%%%%%%%%%%%%%%%%%%%%%%%%%%%%%%%%%%%%%%%%%%

\section{Introduction}
\label{sec:1}

\begin{figure}[t]
 \begin{center}
  \includegraphics[width=0.8\textwidth]{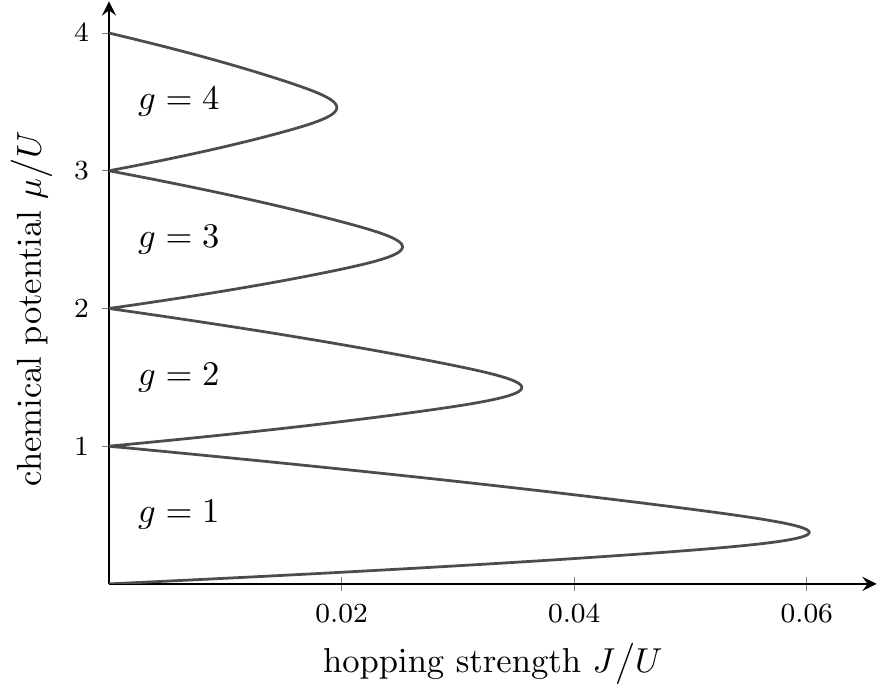}
 \end{center}
 \caption{Phase diagram for the Bose-Hubbard model on a two-dimensional square lattice at zero temperature. Within the lobes located at low $\JU$ the system is in an incompressible Mott state with $g$~particles per site, outside these lobes in a superfluid state. The tips of the lobes represent multicritical points; here the system falls into the universality class of the three-dimensional $XY$~model. This diagram has been computed according to the hypergeometric scheme developed in Ref.~\cite{SandersHolthaus17b}.}      
 \label{F_1}
\end{figure}

Continuous phase transitions are often described by Landau's approach~\cite{Landau37,Landau69,LaLifV,ZinnJustin02,AmitMartinMayor05}: Assume that the thermodynamical potential~$\Gamma$ of a given system possesses the form
\begin{equation}
 \Gamma = a_0 + a_2 \psi^2 + a_4 \psi^4 \; , 
\label{eq:GammA}
\end{equation}
where the coefficients $a_0$, $a_2$, $a_4$ depend on a control parameter~$j$, and the system adopts, for each fixed value of $j$, that value $\psimin$ of $\psi$ for which the potential~\eqref{eq:GammA} takes on its minimum. If then $a_4$ is positive and thus guarantees stability, and if one may further neglect the dependence of $a_4$ on~$j$, while $a_2$ crosses zero at some value $j_{\rm c}$, being positive for $j < j_{\rm c}$ and negative for $j > j_{\rm c}$, one finds 
\begin{equation}
 \psimin = 0	\qquad {\rm for} \quad j < j_{\rm c} \; ,
\end{equation}
whereas
\begin{equation}
 \psimin = \left( \frac{-a_2}{2 a_4} \right)^{1/2}	\qquad {\rm for} \quad j > j_{\rm c} \; .
\label{eq:PsimS}
\end{equation}
In particular, if $a_2$ varies linearly with~$j$ according to
\begin{equation}
 a_2(j) = -\alpha \big(j - j_{\rm c} \big)  
\end{equation}
with $\alpha > 0$, one obtains
\begin{equation}
 \psimin = \sqrt{\frac{\alpha}{2 a_4}} \big( j - j_{\rm c})^{1/2}	\qquad {\rm for} \quad j > j_{\rm c} \; .
\label{eq:TrivE}
\end{equation}
Thus, $\psimin$ serves as an order parameter of the transition, emerging with the mean-field exponent $\beta = 1/2$ at the transition point $j_{\rm c}$.\\
In the present work we extend this basic scenario such that it captures the quantum phase transition from a Mott insulator to a superfluid in the pure two-dimensional Bose-Hubbard model at zero temperature. The Bose-Hubbard model is a paradigmatically simple lattice model of many-particle physics, involving spinless nonrelativistic Bose particles which move on a $d$-dimensional lattice of arbitrary geometry~\cite{GerschKnollmann63,FisherEtAl89,TeichmannEtAl10}. Neighboring lattice sites are connected by a tunneling link of strength~$J$, and two particles occupying the same site repel each other with energy~$U$; the dimensionless ratio~$\JU$ then plays the role of the control parameter~$j$. The system is supposed to be open; its particle content being regulated by a chemical potential~$\mu$. The phase diagram resulting for a two-dimensional square lattice in the $\JU$-$\muU$-plane is shown in Fig.~\ref{F_1}; the corresponding diagrams for triangular or hexagonal lattices are available in the literature~\cite{TeichmannEtAl10}. Within the so-called Mott lobes confined at low $\JU$ between successive integer values $g-1$ and $g$ of the scaled chemical potential $\muU$ the system is in an incompressible Mott state with $g$~particles per site; when increasing $\JU$ at fixed $\muU$ it enters the superfluid phase at the phase boundary $\Jcrit$. This quantum phase transition has been studied in quantitative detail by quite a number of authors with various  methods~\cite{FreericksMonien96,CapogrossoSansoneEtAl07,PolakKopec07,SantosPelster09,TeichmannEtAl09,FreericksEtAl09,HeilvonderLinden12}; it reflects the competition between the lowering of the kinetic energy with increasing delocalization, and the lowering of the repulsion energy with increasing localization of the particles.\\
It is known from the scaling theory of Fisher \etal~\cite{FisherEtAl89} that generally this Mott insulator-to-superfluid transition is mean field-like in character, but with the exception of the multicritical points with particle-hole symmetry at the tips of the Mott lobes, where the transition takes place at fixed density corresponding to an integer filling factor~$g$. At these special points the transition shown by the $d$-dimensional Bose-Hubbard model falls into the universality class of the $(d+1)$-dimensional $XY$~model. Thus, the case $d = 2$ is of particular interest, since it leads to the three-dimensional $XY$ universality class, which also covers the intensely
studied lambda-transition undergone by liquid helium~\cite{LipaEtAl03}.\\
When trying to reconcile this existing knowledge with an approach based on an effective potential~\eqref{eq:GammA}, one faces several questions: How do the Landau coefficients $a_{2k}$, which now also depend, besides the control parameter $\JU$, on the scaled chemical potential $\muU$, manage to switch from ``mean field-like'' to ``multicritical'' upon variation of $\muU$? How does one obtain nontrivial critical exponents from this approach, as opposed to the trivial exponent $\beta = 1/2$ showing up in equation~\eqref{eq:TrivE}? What effort would be required to compute these critical exponents along this line with sizeable accuracy? These are the questions we address in the present work, which constitutes a clarification and significant extension of our previous brief communication~\cite{SandersEtAl15}.\\
Traditionally, the calculation of critical exponents is performed within the framework of the renormalization group (RG) theory~\cite{PelissettoVicari02,ZinnJustin02,AmitMartinMayor05,RanconDupuis11}, having produced fairly precise data. Thus, we do not primarily aim at improving the numerical accuracy of known critical exponents. Instead, we intend to establish a novel bootstrap procedure for computing critical exponents which does not make use of RG theory, and therefore might lead to additional insight. The key input into our analysis are the correlation functions which have been obtained in the accompanying Ref.~\cite{SandersHolthaus17b}, referred to as paper~II in the following. In that more technical paper~II we have investigated the analytic continuation of divergent strong-coupling perturbation series for the Bose-Hubbard model by means of generalized hypergeometric functions ${_{q+1}F_q}$, in comparison with the more familiar Shanks transformation and Pad\'e approximation methods, and have found hypergeometric analytic continuation to be particularly well-suited for characterizing the divergence of the correlation functions at the transition points. Nonetheless, the present paper can be read independently from paper~II, since here we require only certain results obtained therein, while detailed working knowledge of the hypergeometric continuation technique as such is not necessary.\\
We proceed as follows: In Sec.~\ref{sec:2} we briefly recapitulate the formal derivation of the appropriate effective potential for the Bose-Hubbard model~\cite{SantosPelster09,TeichmannEtAl09}, and state the required relations between the Landau coefficients and the correlation functions. In the central Sec.~\ref{sec:3} we then show how to evaluate the critical exponent of the order parameter. In view of the existing accurate reference data, this puts hypergeometric continuation to a truly hard, meaningful test. In Sec.~\ref{sec:4} we discuss a property that characterizes the Landau coefficients at
the multicritical points. Finally, the discussion led in Sec.~\ref{sec:5} concludes our investigation. As an interesting conceptual insight gained from our analysis, we find that it may not always suffice to terminate the effective potential after the fourth-order term, as done in the time-honored paradigm~\eqref{eq:GammA}; rather, for extracting the order-parameter exponent describing the Mott insulator-to-superfluid transition one also has to resort to the Landau coefficient~$a_6$. While we do deliberately restrict ourselves to the two-dimensional Bose-Hubbard model for the sake of definiteness, it stands to reason that our methods are also applicable to further systems.
\section{The effective potential for the Bose-Hubbard model}
\label{sec:2}
The Bose-Hubbard model is formulated in terms of operators $\widehat{b}_i^{\dagger}$ and $\widehat{b}_i^{\phantom \dagger}$ which create and annihilate, respectively, a Bose particle at the $i$th lattice site~\cite{GerschKnollmann63,FisherEtAl89}. Thus, they obey the usual commutation relations 
\begin{equation}
 [\widehat{b}_i^{\phantom \dagger}, \widehat{b}_j^{\dagger}] = \delta_{ij} \; ,
\end{equation}
and the local number operators
\begin{equation}
 \widehat{n}_i = \widehat{b}_i^{\dagger}\widehat{b}_i^{\phantom \dagger}
\end{equation}
yield the number of particles occupying the $i$th site. Employing the pair repulsion energy~$U$ as the energy scale of reference, the model Hamiltonian is written in dimensionless form as
\begin{equation}
 \widehat{H}_{\rm BH}
  = \frac{1}{2} \sum_{i} \widehat{n}_i(\widehat{n}_i-1)
  - \muU \sum_{i} \widehat{n}_i
  - \JU \sum_{\langle i,j \rangle} \, \widehat{b}_i^{\dagger}\widehat{b}_j^{\phantom \dagger} \; ,  	
\label{eq:BaseH}
\end{equation}
where the first two sums extend over the entire lattice, and the symbol $\langle i,j \rangle$ is meant to indicate that the third sum ranges over all pairs of neighboring sites $i$ and $j$. Hence, the first term on the right-hand side gives the total repulsion energy, the second specifies the interaction with the given chemical potential, and the third corresponds to the kinetic energy of the particles. As usual in field theory, we couple this system~\eqref{eq:BaseH} to external sources and drains which we choose to be spatially uniform with strength $\eta$, giving the extended system     	
\begin{equation}
 \widehat{H}
  = \widehat{H}_{\rm BH} 
  + \sum_i \eta \left( \widehat{b}_i^{\dagger}
  + \widehat{b}_i^{\phantom \dagger} \right) \; .
\label{eq:FullH}
\end{equation}
Without loss of generality we have taken $\eta$ to be real, since any phase could be removed by an appropriate redefinition of $\widehat{b}_i^{\dagger}$ and $\widehat{b}_i^{\phantom \dagger}$. The key quantity of interest 
for the theoretical description of this model at zero temperature now is the intensive energy landscape ${\mathcal E}(\muU, \JU, \eta) = \langle \widehat H \rangle / M$, where $M$ denotes the total number of lattice sites, which is assumed to be so large that finite-size effects do not matter, and the expectation value is taken with respect to the ground state of the extended system~\eqref{eq:FullH} which, in contrast to the basic system~\eqref{eq:BaseH}, does not conserve the number of particles. Since this ground state energy is an even function of the source strength~$\eta$, we expand it in the form
\begin{equation}
 {\mathcal E}(\muU, \JU, \eta) = e_0(\muU , \JU) + \sum_{k=1}^\infty c_{2k}(\muU, \JU) \, \eta^{2k} , 
\label{eq:SeriE}
\end{equation}
where the coefficients $c_{2k}(\muU,\JU)$ represent $k$-particle correlation functions. In the accompanying paper~II we have shown how to evaluate these correlation functions by means of a combination of high-order perturbation theory and hypergeometric analytic continuation. In particular, we have studied the one-particle correlation function $c_2$ and the two-particle correlation function $c_4$: When approaching the phase boundary from within a Mott lobe by varying the control parameter $\JU$ at fixed chemical potential $\muU$, these functions diverge as    
\begin{equation}
 c_{2k}(\muU, \JU) \sim \Big( \Jcrit - \JU \Big)^{-\epsilon_{2k}(\muU)}
\label{eq:DivgC}
\end{equation}
with certain positive characteristic exponents $\epsilon_{2k}(\muU)$; we have estimated the exponents $\epsilon_2(\muU)$ and $\epsilon_4(\muU)$ numerically for chemical potentials pertaining to the lowest lobes~\cite{SandersHolthaus17b}.\\
Following standard procedures of field theory, the connection between these correlation functions and Landau's description of phase transitions is made by means of a Legendre transformation~\cite{ZinnJustin02,AmitMartinMayor05,ZiaEtAl09}: An effective potential $\Gamma$ is obtained as the Legendre transform of ${\mathcal E}(\muU, \JU, \eta)$ with respect to the source strength~$\eta$~\cite{SantosPelster09}. Thus, we introduce a variable $\psi$ according to
\begin{equation}
 \frac{\partial \mathcal E}{\partial \eta} =: 2 \psi(\eta) \; ;
\end{equation}
the Hellmann-Feynman theorem, applied to the extended, particle number non-conserving Hamiltonian~\eqref{eq:FullH} then immediately gives the relation
\begin{equation}
 \psi(\eta) = \langle \widehat{b}_i^{\phantom \dagger} \rangle \; .
\label{eq:ExvlB}
\end{equation}
The series~\eqref{eq:SeriE} now yields the representation
\begin{equation}
 \psi(\eta) = \sum_{k=1}^\infty k \, c_{2k} \, \eta^{2k-1} \; ,
\end{equation}
which, upon inversion, allows one to express the source strength~$\eta$ in terms of its conjugate variable~$\psi$: 
\begin{equation}
 \eta(\psi)
  = \psi\left( \frac{1}{c_2} - \frac{2c_4}{c_2^4} \psi^2
  + \left( \frac{12 c_4^2}{c_2^7} - \frac{3 c_6}{c_2^6} \right) \psi^4
  + {\mathcal O}(\psi^6) \right) \; .
\end{equation}
Performing the Legendre transformation according to the prescription
\begin{equation}\label{eq:definition_of_the_Legendre_transformation}
 \Gamma\big(\muU, \JU, \psi\big) = {\mathcal E}\big( \muU, \JU, \eta(\psi) \big) - 2\psi \, \eta(\psi) \; ,
\end{equation}
we then obtain the desired effective potential
\begin{eqnarray}
 \Gamma
  & = & e_0 - \frac{1}{c_2} \psi^2 + \frac{c_4}{c_2^4}\psi^4 + \left( \frac{c_6}{c_2^6} - \frac{4 c_4^2}{c_2^7} \right) \psi^6 + {\mathcal O}(\psi^8) \nonumber \\
  & =: & e_0 + a_2 \psi^2 + a_4 \psi^4 + a_6 \psi^6 + {\mathcal O}(\psi^8) \; , \label{eq:EffcP}
\end{eqnarray}
where $e_0 = e_0(\muU, \JU)$ is the intensive ground-state energy of the basic system~\eqref{eq:BaseH}, and the Landau coefficients $a_{2k} = a_{2k}(\muU, \JU)$ ($k = 1,2,3,\ldots$) emerge as certain combinations of the correlation functions. In field-theoretic jargon, these Landau coefficients represent one-particle irreducible (1PI) vertex functions~\cite{ZinnJustin02,AmitMartinMayor05}.\\
Now we are in a situation analogous to the one considered in the Introduction: Since $\eta$ and $\psi$ are Legendre-conjugated variables, we have the identity~\cite{ZiaEtAl09}
\begin{equation}
 \frac{\partial\Gamma}{\partial\psi} = -2\eta \; ;
\end{equation}
since the actual system of interest~\eqref{eq:BaseH} is recovered by setting $\eta = 0$, it corresponds to the stable stationary points $\psimin$ of $\Gamma$. In accordance with equation~\eqref{eq:ExvlB} the Mott insulating phase is characterized by $\psimin = 0$, whereas a nonvanishing value $\psimin \neq 0$ indicates the presence of a superfluid phase, so that $\psimin$ constitutes a \emph{bona fide} order parameter of the Mott insulator-to-superfluid transition. However, we are not free to make convenient assumptions concerning the dependence of the Landau coefficients on the control parameter $\JU$ and the scaled chemical potential $\muU$, but rather have to respect the above connections between the Landau coefficients and the correlation functions determined in paper~II.
\section{Evaluation of the critical exponent}
\label{sec:3}
\begin{figure}[t]
\begin{center}
\includegraphics[width=0.8\textwidth]{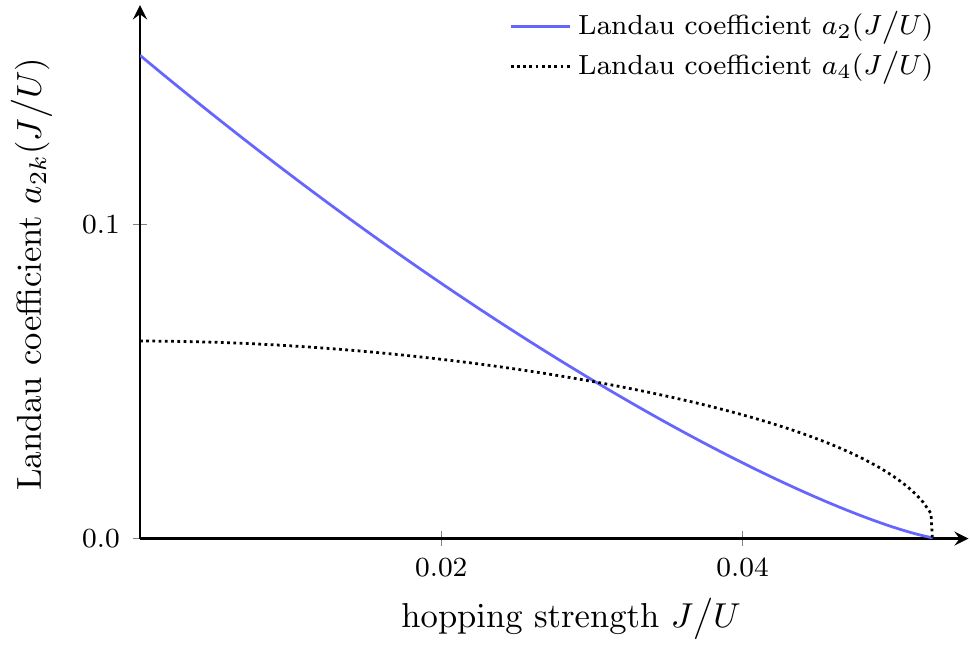}
\end{center}
\caption{(Color online) Behavior of the Landau coefficients $a_2(\JU)$ and $a_4(\JU)$ for $\muU = 0.2652$, as computed from the correlation functions $c_2$ and $c_4$ obtained by hypergeometric analytic continuation based on $_2F_1$ in Ref.~\cite{SandersHolthaus17b}.}
\label{F_2}
\end{figure}

\begin{figure}[t]
\begin{center}
\includegraphics[width = 0.8\textwidth]{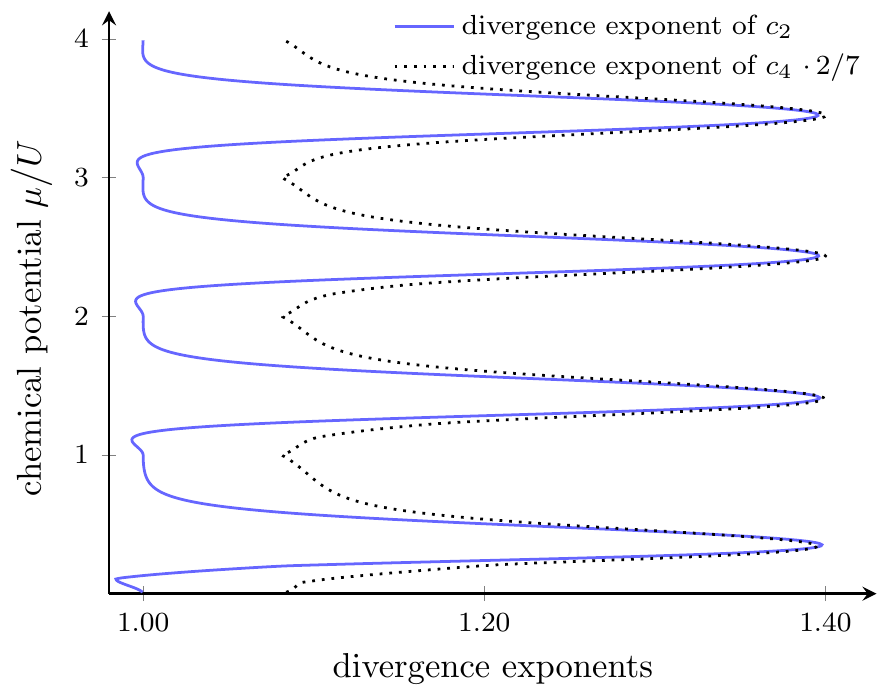}
\end{center}
\caption{(Color online) Divergence exponent $\epsilon_2(\muU)$ of the one-particle correlation function $c_2$ (full line), compared to $2/7$ times the divergence exponent $\epsilon_4(\muU)$ of the two-particle correlation function $c_4$ (dotted line), as computed by hypergeometric analytic continuation based on $_2F_1$. Observe that within the numerical accuracy achieved here one has $\epsilon_4 \ge 7/2 \cdot \epsilon_2$, with equality conjectured at the tips of the lobes.}
\label{F_3}
\end{figure}

The purpose of this section is to investigate the exponents~$\beta = \beta(\muU)$ which govern the emergence of the order parameter according to
\begin{equation}
	\psimin \sim \Big( \JU - \Jcrit \Big)^{\beta(\muU)}
\end{equation}
when $\JU$ is increased at fixed $\muU$ beyond the respective transition point $\Jcrit$. In particular, we will evaluate the exponent $\betacrit$ which belongs to the multicritical points at the tips of the Mott lobes shown in Fig.~\ref{F_1}, where we do expect numerical agreement with the critical exponent $\betacrit = 0.3485(2)$ characterizing the three-dimensional $XY$ universality class~\cite{CampostriniEtAl01}.\\
In paper~II the correlation functions $c_2(\muU,\JU)$ and $c_4(\muU,\JU)$ have been obtained by fitting their strong-coupling perturbation series in the Mott-insulator regime, that is, for $\JU < \Jcrit$ to hypergeometric functions ${_{q+1}F_q}$, thereby determining their divergence exponents~\cite{SandersHolthaus17b}. Here we do not utilize the analytically continued hypergeometric functions for $\JU > \Jcrit$. Instead, the following analysis relies on the assumption that the asymptotic relations~\eqref{eq:DivgC}, namely
\begin{equation}
 c_{2k}(\muU, \JU) \sim \Big( \JU - \Jcrit \Big)^{-\epsilon_{2k}(\muU)}
 \label{eq:asympotics_of_c_2k}
\end{equation}
possess the same divergence exponent $\epsilon_{2k}(\muU)$ on \emph{both} sides of the pole, thus allowing us to make the decisive step into the superfluid regime.\\
For the sake of the argument, let us for the moment assume that for certain $\muU$ we may neglect terms of order ${\mathcal O}(\psi^6)$ in the full effective potential~\eqref{eq:EffcP}. This assumption reduces the effective potential to the archetypal form~\eqref{eq:GammA} reviewed in the Introduction. Its minimum $\psimin$ then is given by
\begin{equation}
	\psimin^2 = \frac{-a_2}{2a_4}
	\qquad {\rm for} \quad \JU > \Jcrit \; ,
\label{eq:PsimA}
\end{equation}
in complete analogy to equation~\eqref{eq:PsimS}. In order to evaluate the exponent $\beta$, we combine the relations~\eqref{eq:EffcP} between the Landau coefficients $a_{2k}$ and the $k$-particle correlation functions $c_{2k}$ with their power-law behavior~\eqref{eq:asympotics_of_c_2k} close to the transition point $\Jcrit$, obtaining
\begin{equation}
	-a_2 = \frac{1}{c_2} 
	\sim \Big( \JU - \Jcrit \Big)^{\epsilon_2(\muU)}
\end{equation}
and
\begin{equation}
	\phantom{-}a_4 = \frac{c_4}{c_2^4} 
	\sim \Big( \JU - \Jcrit \Big)^{4\epsilon_2(\muU) - \epsilon_4(\muU)} \; .
\label{eq:critical_behavior_of_a_4}
\end{equation}
According to equation~\eqref{eq:PsimA} the exponent~$\beta$ would then be given by
\begin{equation}
	\beta = \frac{\epsilon_4 - 3\epsilon_2}{2} \; .
\label{eq:TentE}
\end{equation}
We observe that, in contrast to the example reviewed in the Introduction, the relation~\eqref{eq:critical_behavior_of_a_4} allows $a_4$ to vanish at the phase boundary. This is indeed what happens: Figure~\ref{F_2}, which displays the Landau coefficients $a_2$ and $a_4$ for the arbitrary value $\muU = 0.2652$ of the scaled chemical potential, shows that both $a_2$ and $a_4$, when considered as a function of $\JU$, become zero at $\Jcrit$; the same behavior is found for \emph{all} chemical potentials.\\
Therefore, at the phase boundary we are not entitled to neglect terms of order $\mathcal{O}(\psi^6)$, and have at least to consider the effective potential in the form
\begin{equation}
 \Gamma  = e_0 + a_2 \psi^2 + a_4 \psi^4 + a_6 \psi^6 \; ;
 \label{eq:truncated_potential_considering_psi^6}
\end{equation}
the guiding hypothesis now being that $a_6$ adopts positive values in the superfluid regime. Solving the equation $\Gamma'(\psi_{\rm min}) = 0$ for $\psi_{\rm min}$ then gives
\begin{eqnarray}
	\psimin^2
	  &= \frac{-a_4}{3 a_6}\left(1 \pm \sqrt{1 - \frac{3 a_2 a_6}{a_4^2}} \right)
	  \label{eq:PsimE}\\
	  &= \frac{1}{{12 c_4}/{c_2^3} - {3 c_6}/{c_2^2 c_4}}
		\left(1 \pm \sqrt{\frac{3 c_2 c_6}{c_4^2} - 11}\right) \; .
	  \label{eq:PsimF}
\end{eqnarray}
The underlying assumption
\begin{equation}
 0 < a_6 = \frac{c_6}{c_2^6} - \frac{4 c_4^2}{c_2^7}
\end{equation}
directly entails
\begin{equation}
 \frac{3 c_2 c_6}{c_4^2} - 11 > 1 \; ,
\end{equation}
this means that $c_2 c_6 \big/ c_4^2$ does not converge to zero for $\JU \rightarrow \Jcrit$, which in terms of the divergence exponents $\epsilon_{2k}$ implies that
\begin{equation}
 \epsilon_6 \ge 2 \epsilon_4 - \epsilon_2 \; . \label{eq:epRel}
\end{equation}
In order to deduce the exponent $\beta$ from equation~\eqref{eq:PsimF}, we distinguish two cases:\\
i) In case we have a strict inequality $\epsilon_6 > 2 \epsilon_4 - \epsilon_2$, the combination $c_2 c_6 \big/ c_4^2$ diverges for $\JU \rightarrow \Jcrit$, so that
\begin{equation}
 1 \pm \sqrt{\frac{3 c_2 c_6}{c_4^2} - 11} \sim \sqrt{\frac{3c_2 c_6}{c_4^2}} \; .
\end{equation}
With this, equation~\eqref{eq:PsimF} asymptotically reduces to
\begin{eqnarray}
 \psimin^2
  &\sim \frac{1}{{12 c_4}/{c_2^3} - {3 c_6}/{c_2^2 c_4}} \sqrt{\frac{3c_2 c_6}{c_4^2}} \nonumber \\
  &\sim \frac{1}{\sqrt{3}} \frac{1}{\sqrt{{c_6}/{c_2^5}} - \sqrt{{16 c_4^4}/{c_2^7 c_6}}} \; .
\end{eqnarray}
The proposition $\epsilon_6 > 2 \epsilon_4 - \epsilon_2$ implies $5 \epsilon_2 - \epsilon_6 < 7 \epsilon_2 + \epsilon_6 - 4 \epsilon_4$, and consequently
\begin{equation}
 \frac{c_6}{c_2^5} \gg \frac{c_4^4}{c_2^7 c_6} \qquad {\rm for } \quad \JU \rightarrow \Jcrit \; .
\end{equation}
Hence, the exponent $\beta$ is then given by
\begin{equation}
 \beta = \frac{\epsilon_6 - 5 \epsilon_2}{4} > \frac{\epsilon_4 - 3 \epsilon_2}{2} \; .
\end{equation}
ii) On the other hand, if we have the equality $\epsilon_6 = 2 \epsilon_4 - \epsilon_2$, the square root in equation~\eqref{eq:PsimF} is asymptotically constant and the asymptotics of $\psimin^2$ are given by
\begin{equation}
 \psimin^2 \sim \frac{1}{{12 c_4}/{c_2^3} - {3 c_6}/{c_2^2 c_4}} \; .
 \label{eq:asympotics_of_psimin_at_the_tips}
\end{equation}
Here we have $3 \epsilon_2 - \epsilon_4 = 2 \epsilon_2 + \epsilon_4 - \epsilon_6$, so that both terms in the denominator exhibit the same asymptotic behavior. While it seems mathematically feasible that the leading terms in the difference in the denominator cancel each other, we disregard this unlikely possibility. Therefore, the relation~\eqref{eq:asympotics_of_psimin_at_the_tips} yields the exponent
\begin{equation}
 \beta = \frac{\epsilon_6 - 2 \epsilon_2 - \epsilon_4}{2} = \frac{\epsilon_4 - 3 \epsilon_2}{2} \; .
 \label{eq:RelaI}
\end{equation}
In summary, if terms of order $\mathcal{O}(\psi^8)$ can be neglected in the effective potential~\eqref{eq:EffcP}, the exponent $\beta$ is bounded by
\begin{equation}
 \beta \ge \frac{\epsilon_4 - 3 \epsilon_2}{2} \; ;
 \label{eq:bound_for_beta}
\end{equation}
this bound becomes sharp if the relation~\eqref{eq:epRel} is an equality. Remarkably, the bound equals the previous expression~\eqref{eq:TentE}, which had been deduced from the incorrect proposition that $a_4 > 0$ at the phase transition.\\
We still have to check the current proposition $a_6 > 0$, which is the basis of the result~\eqref{eq:bound_for_beta}.
To do so in full mathematical detail, we have to evaluate the expression
\begin{equation}
 a_6 = \frac{c_6}{c_2^6} - \frac{4c_4^2}{c_2^7} \; ,
\end{equation}
but we lack reliable data\footnote{The perturbative evaluation of $c_6$ requires three creation and three annihilation processes, effectively reducing the number of tunneling events that can still be handled numerically \cite{SandersHolthaus17b}.} for the three-particle correlation function $c_6$, so that we are restricted to the investigation of the term ${c_4^2}/{c_2^7}$. Exemplarily, we again inspect the hypergeometric fits to $c_2$ and $c_4$ at $\muU = 0.2652$, the value already considered in Fig.~\ref{F_2}, and state their divergence exponents $\epsilon_2 = 1.281$ and $\epsilon_4 = 4.621$, respectively. We note that $2 \cdot \epsilon_4 = 9.241 > 8.967 = 7 \cdot \epsilon_2$, signaling that ${c_4^2}/{c_2^7} \sim (\JU - \Jcrit)^{7\epsilon_2 - 2\epsilon_4}$ diverges at the phase boundary, strongly suggesting that $a_6$ shares the same behavior.\\
Inspecting the divergence exponents $\epsilon_2$ and $\epsilon_4$ obtained with hypergeometric analytic continuation for $0 \le \muU \le 4$, as displayed in Fig.~\ref{F_3}, we observe that, to within numerical accuracy, $\epsilon_4 \ge 7/2 \cdot \epsilon_2$ in this entire interval.
Based on the particular shape of the curves drawn in Fig.~\ref{F_3}, we surmise that this actually is a strict inequality aside from the tips, whereas
\begin{equation}
 \epsilon_4 = \frac{7}{2} \, \epsilon_2 \qquad {\rm at~the~tips~of~the~lobes}\; .
\label{eq:IdenY}
\end{equation}
Under the assumption of the validity of this equation, the divergence of $c_4^2$ and that of $c_2^7$ cancel each other at the tips, and $c_4^2/c_2^7$ has a finite, non-zero limit at the phase boundary. If we further assume that the ratio  $c_6/c_2^6$, and hence $a_6$, shares the same behavior, we deduce $\epsilon_6 = 6 \epsilon_2 = 7 \epsilon_2 - \epsilon_2$, which, in view of the equality~\eqref{eq:IdenY}, gives $\epsilon_6 = 2 \epsilon_4 - \epsilon_2$. This is precisely the second case~ii) in the above distinction, for which we have derived the equality~\eqref{eq:RelaI}. This leads to a decisive conclusion: Inserting the relation~\eqref{eq:IdenY} between the divergence exponents into this formula~\eqref{eq:RelaI} for the exponent~$\beta$ of the order parameter, we obtain the identity
\begin{equation}
 \betacrit = \frac{\epsilon_2}{4}
\label{eq:FinaL}
\end{equation}
for the critical exponent $\betacrit$ at the tips of the lobes. While the inequality~\eqref{eq:bound_for_beta} is a general result, this equality~\eqref{eq:FinaL} hinges on the observations made in Fig.~\ref{F_3}, and applies to the multicritical points only. We thus arrive at an interesting characterization of the multicritical points: the Landau coefficient $a_6$ diverges for all chemical potentials when $\JU \rightarrow \Jcrit$, with the exception of these points.\\
The key result~\eqref{eq:FinaL}, stating that the critical exponent $\betacrit$ at the tips of the Mott lobes is given by one fourth of the divergence exponent~$\epsilon_2$ of the two-particle correlation function~$c_2$, is amenable to quantitative verification: In Tab.~\ref{T_1} we list the values of $\betacrit$ as obtained from equation~\eqref{eq:FinaL} at the tip of the lowest Mott lobe $g = 1$ by hypergeometric continuation based on ${_{q+1}F_q}$ with~$q$ ranging from $0$ to $4$; the third column states the relative deviation of the respective result from the reference value~$\betacrit = 0.3485(2)$ which has been derived from the $\phi^4$ lattice model and the dynamically diluted $XY$-model by combining Monte Carlo simulations based on finite-size scaling methods, and high-temperature expansions~\cite{CampostriniEtAl01}. Evidently, the agreement is quite good.\\
\begin{table}
\begin{center}
\begin{tabular}{c|c|c}
 fit function~ 	& $\betacrit$ 	& ~relative deviation	\\
\hline
 $_1F_0$	&~0.3511~		&  \phantom{-}0.76\%	\\
 $_2F_1$	& 0.3475		&           - 0.30\%	\\
 $_3F_2$	& 0.3441		&           - 1.27\%	\\
 $_4F_3$	& 0.3459		&           - 0.76\%	\\
 $_5F_4$	& 0.3478		&           - 0.18\%	\\
\end{tabular}
\end{center}
\caption{Comparison of the critical exponent as obtained at the tip of the first Mott lobe $g = 1$ by hypergeometric analytic continuation based on $_{q+1}F_q$ with the value $\betacrit = 0.3485(2)$ expected for the three-dimensional $XY$ universality class.}
\label{T_1}
\end{table}
Speculating further that ${_2F_1}$ might yield the most accurate numerical estimates, representing a good compromise between flexibility, as provided by the number of fitting parameters, and the number of available input data still accessible to high-order perturbation theory, we also present estimates for $\betacrit$ computed with ${_2F_1}$ for the lowest four Mott lobes in Tab.~\ref{T_2}. Based on these data, we cautiously state our final result $\betacrit = 0.348(1)$.\\
\begin{table}
\begin{center}
\begin{tabular}{c|c|c}
 lobe index~$g$~	& $\betacrit$	& ~relative deviation	\\
 \hline
 1			&~0.3475~		&          - 0.30\%	\\
 2			& 0.3483		&          - 0.06\%	\\
 3			& 0.3485		& \phantom{-}0.00\%	\\
 4			& 0.3489		& \phantom{-}0.12\%	\\
\end{tabular}
\end{center}
\caption{Comparison of the critical exponent as obtained by hypergeometric analytic continuation with ${_2F_1}$ at the tip at the lowest four Mott lobes, with the value $\betacrit = 0.3485(2)$ expected for the three-dimensional $XY$ universality class.}
\label{T_2}
\end{table}
\section{Characterization of the critical effective potential}
\label{sec:4}
The previous observation that the Landau coefficient $a_6$ diverges for all chemical potentials at the phase boundary, except at the multicritical points, necessitates further investigations. For motivation, let us once again consider the truncated potential~\eqref{eq:GammA}, which yields the necessary condition
\begin{equation}
 0 \stackrel{!}{=} \frac{\partial \Gamma}{\partial \psi} = 2 a_2 \psi + 4 a_4 \psi^3
\end{equation}
for its minimum $\psimin$. This immediately implies that
\begin{equation}
 \frac{a_4 \psimin^4}{a_2 \psimin^2} = -\frac{1}{2} \; ,
\end{equation}
independent of $\JU$. Consequently, the quadratic and the quartic term, that is, $a_2 \psimin^2$ and $a_4 \psimin^4$, have the same asymptotic behavior for $\JU \rightarrow \Jcrit$.\\
We now return to the full potential~\eqref{eq:EffcP}. Investing only the relations $a_2 = -1/c_2$ and $a_4 = c_4/c_2^4$, we deduce
\begin{equation}
 \frac{a_4 \psimin^4}{a_2 \psimin^2} = -\frac{c_4}{c_2^3} \psimin^2 \sim \Big(\JU - \Jcrit\Big)^{2 \beta - (\epsilon_4 - 3 \epsilon_2)} \; .
\end{equation}
Therefore, both terms $a_2 \psimin^2$ and $a_4 \psimin^4$ have the same asymptotic behavior for $\JU \rightarrow \Jcrit$ if and only if
\begin{equation}\label{eq:formula_for_the_critical_exponent}
 \beta = \frac{\epsilon_4 - 3 \epsilon_2}{2} \; ,
\end{equation}
which is our previous equality~\eqref{eq:RelaI}, valid at the lobe tips. Thus, our formula for the critical exponent at the tips of the Mott lobes implies that the quadratic and the quartic term display the same asymptotic behavior, and vice versa.\\
Going one step further, we observe that
\begin{equation}
 \frac{a_6 \psimin^6}{a_4 \psimin^4} = \frac{{c_6}/{c_2^6} - 4 {c_4^2}/{c_2^7}}{{c_4}/{c_2^4}} \psimin^2 = \left(\frac{c_6}{c_2^2c_4} - \frac{4c_4}{c_2^3}\right) \psimin^2 \; .
\end{equation}
If we now utilize the relation $\epsilon_6 = 2 \epsilon_4 - \epsilon_2$, as strongly supported by our numerical findings at the tips of the lobes, both addends share the same asymptotic behavior for $\JU \rightarrow \Jcrit$. Therefore, we meet the same pattern: Equation~\eqref{eq:formula_for_the_critical_exponent} is equivalent to $a_4 \psimin^4$ and $a_6 \psimin^6$ showing the same behavior.\\
This finding appears to hold in all orders. In general, as a consequence of the Legendre transformation~\eqref{eq:definition_of_the_Legendre_transformation} the Landau coefficient $a_{2k}$ contains an addend which depends on $c_2$ and $c_4$ only, so that
\begin{equation}
 a_{2k} = -(-2)^{k-1} \cdot k! \cdot \frac{c_4^{k-1}}{c_2^{3k -2}} + \cdots \; .
\end{equation}
Generalizing our previous arguments, we conjecture that any two terms $a_{2k} \psimin^{2k}$ and $a_{2l} \psimin^{2l}$ of the effective potential~\eqref{eq:EffcP} exhibit the same asymptotics for $\JU \rightarrow \Jcrit$ if and only if equation~\eqref{eq:formula_for_the_critical_exponent} holds. This observation also resolves an apparent contradiction: Equation~\eqref{eq:formula_for_the_critical_exponent} has been obtained from the truncated potential~\eqref{eq:truncated_potential_considering_psi^6}, although such a truncation is not valid when all terms of the full potential are of the same magnitude. However, equation~\eqref{eq:formula_for_the_critical_exponent} reflects a system property at the multicritical points which is not affected by the truncation, which is why it even has emerged, albeit as the result of an oversimplified reasoning, in equation~\eqref{eq:TentE}.\\
To conclude, there is strong evidence that at the multicritical points corresponding to the tips of the Mott lobes the divergence exponents $\epsilon_{2k}$ are not independent of each other, but can all be related to $\epsilon_2$, as exemplified by our relations $\epsilon_4 = {7}/{2} \cdot \epsilon_2$ and $\epsilon_6 = 2 \epsilon_4 - \epsilon_2 = 6 \epsilon_2$. This is tantamount to the observation that all terms in the effective potential~\eqref{eq:EffcP} display the same asymptotic behavior, and allows us to express the critical exponent $\betacrit$ in terms of $\epsilon_2$ alone, cf. equation~\eqref{eq:IdenY}.
\section{Discussion}
\label{sec:5}  
In this paper we have established a connection between the divergence exponents $\epsilon_{2k}$ of the $k$-particle correlation functions $c_{2k}$ pertaining to the two-dimensional Bose-Hubbard model, as defined by equations~\eqref{eq:SeriE} and \eqref{eq:DivgC}, and the critical exponent of the order parameter of the Mott insulator-to-superfluid transition. This allows us to take advantage of the fact that the divergence exponents $\epsilon_2$ and $\epsilon_4$ can be computed numerically with tolerable effort for any value of the scaled chemical potential $\muU$. This is achieved by means of hypergeometric analytic continuation of the strong-coupling perturbation series of $c_{2}$ and $c_{4}$, respectively, as demonstrated in detail in paper~II~\cite{SandersHolthaus17b}.\\
Under the assumption that the effective potential~\eqref{eq:EffcP} can be truncated after the sixth order term, requiring the Landau coefficient $a_6$ to be positive, we have derived the lower bound
\begin{equation*}
\beta(\muU) \ge \frac{\epsilon_4(\muU) - 3\epsilon_2(\muU)}{4}
\end{equation*}
for the exponent $\beta(\muU)$ with which the order parameter emerges at the Mott insulator-to-superfluid transition.\\
For all chemical potentials except those marking the multicritical points the transition is expected to be mean field-like~\cite{FisherEtAl89}; the bound then is well compatible with the mean-field exponent $\beta_{\rm mf} = 1/2$. At multicriticality, that is, at the lobe tips the bound becomes sharp, and actually yields, to within the numerical accuracy achieved here, the critical exponent $\betacrit$ of the three-dimensional $XY$ class. Moreover, at multicriticality the divergence exponents are no longer independent of each other, but can all be expressed in terms of $\epsilon_2$. Utilizing the ``multicritical'' equality $\epsilon_4 = 7/2 \cdot \epsilon_2$, deduced numerically from Fig.~\ref{F_3}, we arrive at the identity $\betacrit = \epsilon_2/4$, checked to sub-percent accuracy in Tabs.~\ref{T_1} and~\ref{T_2}.\\
The numerical accuracy of our present estimate $\betacrit = 0.348(1)$ does not yet match that of the elaborate Monte Carlo value reported in Ref.~\cite{CampostriniEtAl01} for the $XY$ model, i.e., $\betacrit = 0.3485(2)$. However, both results are well compatible with each other, providing an impressive example of universality in phase transitions.\\
Our approach to critical exponents is essentially self-contained, and comparatively straightforward.
Along the lines pioneered in this paper, hypergeometric continuation for evaluating divergence exponents may provide critical exponents for wide classes of models, thus shedding further light on the universality hypothesis of statistical physics.
%
%%%%%%%%%%%%%%%%%%%%%%%%%
% Acknowledgements
%%%%%%%%%%%%%%%%%%%%%%%%%
%
\ack
We acknowledge CPU time granted to us on the HPC cluster HERO, located at the University of Oldenburg and funded by the DFG through its Major Research Instrumentation Programme (INST 184/108-1 FUGG), and by the Ministry of Science and Culture (MWK) of the Lower Saxony State.
%
%
%%%%%%%%%%%%%%%%%%%%%%%%%
%   References 
%%%%%%%%%%%%%%%%%%%%%%%%%
%
\section*{References}


\begin{thebibliography}{99}
\bibitem{Landau37}
	L. D. Landau,
	Zh. Eksp. Teor. Fiz. {\bf 7}, 19 (1937).

\bibitem{Landau69} 
	L. D. Landau,
	{\em Collected Papers\/}
	(Nauka, Moskow, 1969) Vol.~1, p.~234.

\bibitem{LaLifV} 
	L. D. Landau and E. M. Lifshitz,
	{\em Course of Theoretical Physics, Volume 5: Statistical Physics,
	Part~1\/}
	(Elsevier Butterworth-Heinemann, Oxford, 1980).

\bibitem{ZinnJustin02}
	J. Zinn-Justin,
	{\em Quantum Field Theory and Critical Phenomena\/}, fourth edition
	(Clarendon Press, Oxford, 2002).

\bibitem{AmitMartinMayor05}
	D. J. Amit and V. Mart\'{i}n-Mayor,
	{\em Field Theory, the Renormalization Group, and Critical Phenomena:
	Graphs to Computers\/}, third edition
	(World Scientific, Singapore, 2005).

\bibitem{GerschKnollmann63}
	H. A. Gersch and G. C. Knollmann,
	Phys. Rev. {\bf 129}, 959 (1963).

\bibitem{FisherEtAl89}
	M. P. A. Fisher, P. B. Weichman, G. Grinstein, and D. S. Fisher,
	Phys. Rev. B {\bf 40}, 546 (1989).

\bibitem{TeichmannEtAl10}
	N. Teichmann, D. Hinrichs, and M. Holthaus,
	EPL {\bf 91}, 10004 (2010).

\bibitem{FreericksMonien96}
	J. K. Freericks and H. Monien,
	Phys. Rev. B {\bf 53}, 2691 (1996).

\bibitem{CapogrossoSansoneEtAl07}
	B. Capogrosso-Sansone, N. V. Prokof'ev, and B. V. Sistunov,
	Phys. Rev. B {\bf 75}, 134301 (2007).

\bibitem{PolakKopec07}
	T. P. Polak and T. K. Kope\'{c},
	Phys. Rev. B {\bf 76}, 094503 (2007).

\bibitem{SantosPelster09}
	F. E. A. dos Santos and A. Pelster,
	Phys. Rev. A {\bf 79}, 013614 (2009).

\bibitem{TeichmannEtAl09}
	N. Teichmann, D. Hinrichs, M. Holthaus, and A. Eckardt,
	Phys. Rev. B {\bf 79}, 100503(R) (2009).

\bibitem{FreericksEtAl09}
	J. K. Freericks, H. R. Krishnamurthy, Y. Kato, N. Kawashima,
	and N. Trivedi,
	Phys. Rev. A {\bf 79}, 053631 (2009).

\bibitem{HeilvonderLinden12}
	C. Heil and W. von der Linden,
	J. Phys.: Condens. Matter {\bf 24}, 295601 (2012).

\bibitem{LipaEtAl03}
	J. A. Lipa, J. A. Nissen, D. A. Stricker, D. R. Swanson,
	and T. C. P. Chui,
	Phys. Rev. B {\bf 68}, 174518 (2003).

\bibitem{SandersEtAl15}
	S. Sanders, C. Heinisch,  and M. Holthaus,
	EPL {\bf 111}, 20002 (2015).

\bibitem{PelissettoVicari02}
	A. Pelissetto and E. Vicari,
	Phys. Rep. {\bf 368}, 549 (2002).

\bibitem{RanconDupuis11}
	A Ran\c{c}on and N. Dupuis,
	Phys. Rev. B {\bf 84}, 174513 (2011).

\bibitem{SandersHolthaus17b}
	S. Sanders and M. Holthaus,
	J. Phys. A: Math. Theor., in press https://doi.org/10.1088/1751-8121/aa8f01

\bibitem{ZiaEtAl09}
	R. K. P. Zia, E. F. Redish, and S. R. McKay,
	Am. J. Phys. {\bf 77}, 614 (2009).

\bibitem{CampostriniEtAl01}
	M. Campostrini, M. Hasenbusch, A. Pelissetto, P. Rossi, and E. Vicari,
	Phys. Rev. B {\bf 63}, 214503 (2001).

\end{thebibliography}
\end{document}